\documentclass[10pt]{article}
\usepackage{jheppub}
\usepackage{axodraw2}
\allowdisplaybreaks 

\newcommand{\ba}{\begin{eqnarray}}
\newcommand{\ea}{\end{eqnarray}}
\newcommand{\af}{a_f}
\newcommand{\cf}{c_f}
\newcommand{\nf}{n_f}
\newcommand{\str}{{\rm sTr}}
\newcommand{\la}[1]{\label{#1}}
\newcommand{\fig}{figure~}
\newcommand{\eq}{eq.~}
\newcommand{\se}{section~}
\newcommand{\secs}{sections~}
\newcommand{\app}{app.~}
\newcommand{\eqs}{eqs.~}
\newcommand{\nr}[1]{(\ref{#1})}

\newcommand{\ep}{\varepsilon}
\newcommand{\CA}{C_{\mathrm{A}}}
\newcommand{\CF}{C_{\mathrm{F}}}
\newcommand{\NA}{N_{\mathrm{A}}}
\newcommand{\NF}{N_{\mathrm{F}}}
\newcommand{\TF}{T_{\mathrm{F}}}
\newcommand{\Nf}{N_{\mathrm{f}}}
\newcommand{\order}[1]{{\cal O}(#1)}
\newcommand{\msbar}{{\overline{\mbox{\rm{MS}}}}}
\newcommand{\code}[1]{{\tt #1}}
\newcommand{\PolyGamma}{\psi}

\newcommand{\gammaQQ}{\gamma_2}

\newcommand{\picbj}[1]{\;\parbox[c]{60pt}{\begin{picture}(60,40)(0,0)
\SetWidth{1.0}\SetScale{1.0} #1 \end{picture}}\;}
\newcommand{\picbjj}[1]{\parbox[c]{0pt}{\begin{picture}(0,40)(63,0)
\SetWidth{1.0}\SetScale{1.0} #1 \end{picture}}}
\newcommand{\sbx}{\scalebox{0.825}}
\newcommand{\defDiag}[2]{\expandafter\newcommand%
  \csname diag-#1\endcsname{#2}}
\newcommand{\diag}[1]{\csname diag-#1\endcsname}
\def\Asc(#1,#2)(#3,#4,#5){\CArc(#1,#2)(#3,#4,#5)}
\def\Lsc(#1,#2)(#3,#4){\Line(#1,#2)(#3,#4)}
\defDiag{31740}{\sbx{\picbj{
\Lsc(20,0)(40,0)
\Lsc(20,40)(40,40)
\Asc(20,20)(20,-180,-90)
\Asc(40,20)(20,-90,90)
\Lsc(20,0)(20,20)
\Lsc(40,20)(40,0)
\Lsc(20,20)(20,40)
\Lsc(40,20)(40,40)
\Asc(20,20)(20,90,180)
\Lsc(20,13)(40,13)
\Lsc(20,27)(40,27)}}}
\defDiag{30527}{\sbx{\picbj{
\Asc(20,20)(20,90,270)
\Asc(40,20)(20,-90,90) 
\Lsc(40,40)(20,40) 
\Lsc(20,0)(40,0) 
\Lsc(20,0)(20,20)
\Lsc(20,20)(40,40) 
\Lsc(20,40)(28,33) 
\Lsc(33,28)(40,20) 
\Lsc(40,20)(40,0)
\Lsc(20,20)(40,20)
\Lsc(30,0)(30,20)}}}
\defDiag{31740.1.3}{\diag{31740}\sbx{\picbjj{\Vertex(58.5,27){2.5}\Vertex(58.5,13){2.5}}}}
\defDiag{30527.1.3}{\diag{30527}\sbx{\picbjj{\Vertex(1.5,27){2.5}\Vertex(1.5,13){2.5}}}}
\defDiag{30527.1.2}{\diag{30527}\sbx{\picbjj{\Vertex(0,20){2.5}}}}
\defDiag{30527.6.3}{\diag{30527}\sbx{\picbjj{\Vertex(20,7){2.5}\Vertex(20,13){2.5}}}}
\defDiag{30527.1.4}{\diag{30527}\sbx{\picbjj{\Vertex(4,32){2.5}\Vertex(0,20){2.5}\Vertex(4,8){2.5}}}}
\newcommand{\TLfig}[1]{{\begin{array}{c}\diag{#1}\\[2ex] I_{#1}\end{array}}}

\title{Five-loop quark mass and field anomalous dimensions for a general gauge group}

\preprint{BI-TP 2016/08\\\mbox{}\hfill DESY 16-245\\\mbox{}\hfill IPPP/16/120}

\author[a]{Thomas Luthe,}
\author[b]{Andreas Maier,}
\author[c]{Peter Marquard}
\author[d]{and York Schr\"oder}

\affiliation[a]{Faculty of Physics, University of Bielefeld, 33501 Bielefeld, Germany}
\affiliation[b]{Institute for Particle Physics Phenomenology, Durham University, Durham, United Kingdom}
\affiliation[c]{Deutsches Elektronen Synchrotron (DESY), Platanenallee 6, Zeuthen, Germany}
\affiliation[d]{Grupo de F\'isica de Altas Energ\'ias, Universidad del B\'io-B\'io, Casilla 447, Chill\'an, Chile}

\emailAdd{tluthe@physik.uni-bielefeld.de}
\emailAdd{andreas.maier@durham.ac.uk}
\emailAdd{peter.marquard@desy.de}
\emailAdd{yschroder@ubiobio.cl}

\arxivnumber{1612.05512}

\abstract{We present analytical five-loop results for the quark mass and quark field anomalous dimensions, for a general gauge group and in the $\msbar$ scheme. We confirm the values known for the gauge group SU(3) from an independent calculation, and find full agreement with results available from large-\/$\Nf$ studies.}

\begin{document}
\maketitle

%
\section{Introduction}
\la{se:intro}

High-precision determinations of Standard Model (SM) parameters are crucially important
for precise predictions for the observables measured in collider physics experiments,
such as currently performed at the LHC, or possibly at a future linear collider. 
A thorough comparison of these predictions with experimental results allows to scrutinize 
the details of the hugely successful SM, and might shed light on possible physics beyond the SM.

In order to pin down the theory's fundamental parameters, such as coupling constants and 
masses, with sufficient precision, the knowledge of higher order perturbative corrections is
required. This encompasses the evaluation of multi-loop Feynman diagrams and Feynman
integrals, for which significant progress has been made in recent years, mainly with respect
to the formulation of advanced algorithms that allow to treat the complexity level met in those
higher-loop integrals.

We focus here on the strong interactions, which are embedded into the SM via Quantum 
Chromodynamics (QCD). The relevant parameters are then the (strong) gauge coupling 
and the quark masses, both of which run with the energy scale according to the renormalization
group (RG) equations. In order to evolve e.g.\ the low-energy value of the coupling constant
(measured with high precision from tau lepton decay) to high energies, the anomalous dimension
of the gauge coupling (the so-called Beta function) is needed, as coefficient in the corresponding
RG equation.
Likewise, a high-order evaluation of the quark mass anomalous dimension gives access
to precise values for e.g.\ charm and bottom quark masses, which are measured at low 
energies (typically a few GeV) but whose uncertainty at the high-energy scale of the Higgs mass $m_H=125$ GeV is important in Higgs decay rates into such quark pairs.
In particular, to match the precision of the known five-loop inclusive decay width of $H\rightarrow q\bar q$ \cite{Baikov:2005rw}, one should evolve the parameters (which are $\alpha_s(\mu)$ and the running quark mass $m_q(\mu)$ with $\mu$ being the renormalization scale) from low energies to $\mu=m_H$ at the same perturbative order, for full consistency and to avoid large logarithms $\ln(\mu^2/m_H^2)$.

Given this clear phenomenological motivation, we will present new results for the quark mass
anomalous dimension here, applying a number of the above-mentioned algorithmic advances. 
While this renormalization constant has been studied previously up to 
five loops in perturbation theory \cite{Tarrach:1980up,Tarasov:1982gk,Larin:1993tq,Vermaseren:1997fq,Chetyrkin:1997dh,Baikov:2014qja}, we generalize it from the gauge group SU(3) to a semi-simple
Lie group. At the same time, we provide a truly independent check on the available SU(3) result, 
since we utilize largely independent methods, as described below.
We will also give results for a related quantity needed to renormalize the quark sector at five loops, 
namely the quark field anomalous dimension in Feynman gauge, again generalizing known SU(3) 
results to a semi-simple Lie group.

The paper is organized as follows. 
We start by explaining our computational setup in \se\ref{se:setup}.
Using some notation defined in \se\ref{se:notation}, we then present and discuss results in \secs \ref{se:gm} and \ref{se:g2}, before concluding in \se\ref{se:conclu}. For convenience, an appendix reproduces large-\/$\Nf$ results from the literature, which we use for consistency checks.

%
\section{Setup}
\la{se:setup}

Let us start by explaining our computational setup, which closely follows the one employed and tested in \cite{Luthe:2016ima}.
We base our highly automated setup on the diagram generator \code{qgraf} \cite{Nogueira:1991ex,Nogueira:2006pq} and various in-house \code{FORM} \cite{Vermaseren:2000nd,Tentyukov:2007mu,Kuipers:2012rf} codes.
After generating the required fermionic 2-point functions, we apply projectors and perform the group algebra with \code{color} \cite{vanRitbergen:1998pn}. To extract the ultraviolet (UV) divergences, we then use the freedom to change the low-momentum behavior and make all propagators massive, which regulates the infrared at the cost of introducing a new counterterm for the unphysical regulator mass. Expanding to sufficient depth in the external momentum \cite{Misiak:1994zw,vanRitbergen:1997va,Chetyrkin:1997fm} and keeping all potentially UV divergent structures results in nullifying the external momentum. The coefficients of this expansion can then be mapped onto a family of fully massive vacuum integrals, which are labelled by 15 indices (corresponding to maximally 12 propagators plus 3 scalar products) at five loops \cite{Luthe:2015ngq}. 
As a next and fairly time-consuming step, we reduce those integrals to a small set of master integrals, powered by our own codes \code{crusher} \cite{crusher} and \code{TIDE} \cite{Luthe:2015ngq}, which are based on integration-by-parts (IBP) identities \cite{Chetyrkin:1981qh} and use a Laporta-type algorithm \cite{Laporta:2001dd} for a systematic integral reduction.

At five loops, we end up with a set of 110 master integrals. These have been evaluated in an $\ep$\/-expansion around $d=4-2\ep$ dimensions in previous works \cite{Luthe:2015ngq,Luthe:2016sya}, using an approach based on IBP reductions and difference equations \cite{Laporta:2001dd} that has been realized in \code{C++} and uses \code{Fermat} \cite{fermat} to perform the polynomial algebra that arises in solving systems of linear equations with large rational coefficients. 
The resulting high-precision numerical results for the coefficients of the $\ep$\/-expansions finally allow us to utilize the integer-relation finding algorithm \code{PSLQ} \cite{MR1489971} to discover the analytic content of some of these numbers, and to find relations between others. As a consequence, we are able to provide all our results given below in analytic form.

\begin{figure}
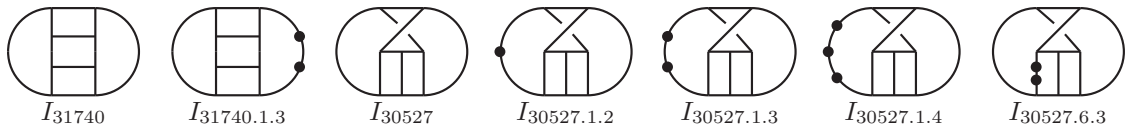

\SetScale{2.0}
\begin{center}
\begin{align*}
\begin{array}{ccccccc}
\TLfig{31740}
&\TLfig{31740.1.3}
&\TLfig{30527}
&\TLfig{30527.1.2}
&\TLfig{30527.1.3}
&\TLfig{30527.1.4}
&\TLfig{30527.6.3}
\end{array}
\end{align*}
\end{center}
\caption{\label{fig:LCgraphs}5-loop master integrals with 12 lines that contribute to \eqs\nr{eq:lc0}-\nr{eq:lc2}. Each line denotes a massive propagator $1/(k^2+m^2)$, and each dot stands for an extra power of the corresponding propagator.}
\end{figure}

\newcommand{\lcA}{\ell_0}
\newcommand{\lcB}{\ell_1}
\newcommand{\lcC}{\ell_2}

As has already been mentioned elsewhere \cite{Luthe:2016ima}, our high-precision evaluation of all 5-loop master integrals has not yet produced results for the 12-line families, see \fig\ref{fig:LCgraphs}. These do not contain divergences in four dimensions, and could therefore be avoided in evaluations of anomalous dimensions. It turns out, however, that with our integral reduction criteria and lexicographic ordering prescription, we do get contributions from these integral classes since they are multiplied by prefactors with spurious poles as $d\rightarrow4$. 
To fix the values of the 12-line master integrals that we need, we first note that in all our results, only three independent linear combinations appear: 
\ba
\label{eq:lc0}
\lcA &=& 689 \,I_{30527.1.1, 5} - 3934 \,I_{30527.1.2, 5} + 5464 \,I_{30527.1.3, 5} + 1152 \,I_{30527.1.4, 5} - 5228 \,I_{30527.6.3, 5} \;,\\
\lcB &=& 689 \,I_{30527.1.1, 6} - 3934 \,I_{30527.1.2, 6} + 5464 \,I_{30527.1.3, 6} + 1152 \,I_{30527.1.4, 6} - 5228 \,I_{30527.6.3, 6} \nonumber\\&&
+ 1968 \,I_{30527.1.1, 5} + 3890 \,I_{30527.1.2, 5} + 3844 \,I_{30527.1.3, 5} - 6912 \,I_{30527.1.4, 5} - 70 \,I_{30527.6.3, 5} \;,\\
\label{eq:lc2}
\lcC &=& 11 \,I_{31740.1.1, 5} - 72 \,I_{31740.1.3, 5} \;.
\ea
Here, each integral $I_{\#,n}$ corresponds to the $\ep^n$\/-coefficient of the respective fully massive master integral of \fig\ref{fig:LCgraphs}, divided by the fifth power of the 1-loop tadpole $J$ for normalization reasons\footnote{$J=\int{\rm d}^dk/(k^2+m^2)\sim\Gamma(1-d/2)$ has a simple pole as $d\rightarrow4$; hence, finite 5-loop terms correspond to $\ep^5$.}.
While it is conceivable that there exists a suitable basis transformation that eliminates these linear combinations altogether from the final results, we have not yet performed a systematic search of such transformations in our integral reduction tables, but opted for other criteria to fix the numerical values of the three linear combinations, with high precision, as we will explain now.

As a crude order-of-magnitude estimate, we have evaluated the set of 12-line integrals via Feynman parametric representations (see e.g.\ \cite{Smirnov:2012gma}) and subsequent (primary) sector decomposition, using the strategy explained in \cite{Roth:1996pd,Binoth:2003ak} and as implemented in \code{FIESTA} \cite{Smirnov:2015mct} as well as own code (see \cite{Luthe:2016spi}). Due to the large prefactors and cancellations in \eqs\nr{eq:lc0}-\nr{eq:lc2}, a 6-digit evaluation results in the estimates 
\ba
\label{eq:LCestimates}
\lcA\approx-7.47(1)\;,\quad 
\lcB\approx-50.6(1)\;,\quad 
\lcC\approx-0.673(1)\;,
\ea 
with 3-digit accuracy.

Turning now to high-precision evaluations,
we first recall that the higher-order $\ep$\/-poles of renormalization constants are completely determined by lower-order coefficients. Checking these, we obtain one consistency condition that allows to fix $\lcA$.
Second, we observe the occurrence of some rank-12 group invariants in another renormalization constant that we have evaluated using the same setup, namely for the ghost-gluon vertex \cite{inPrep}. Using \eq\nr{eq:LCestimates}, to determine their coefficients, we find that they vanish at least to this low accuracy. Taking this zero for granted, we turn the argument around and require e.g.\ the structure $d^{\,444}_{FAA}\,\Nf$ (where the group invariant is in the notation of \cite{vanRitbergen:1998pn}) to be absent from the final result; this gives us another condition, fixing $\lcC$. Third, for fixing the remaining linear combination, we choose to compare our results in the SU(3) limit to the previously known 5-loop results. To be concrete, out of the many possible coefficients we choose the $\nf$ term of $\gamma_m$ as given in \cite{Baikov:2014qja}, giving us one constraint which fixes $\lcB$. 
Along these lines, we obtain numerical values for all three linear combinations with 260 digits, the first 50 of which read
\ba
\lcA &=& -7.4750787021276651819913288152084850401974826928834\dots \;,\\
\lcB &=& -50.563714841071996428539372592222326105092965639946\dots \;,\\
\lcC &=& -0.67332086607447050046759024439428336720209195028580\dots \;,
\ea
and which can be seen to be consistent with our low-precision estimates of \eq\nr{eq:LCestimates}
that had been obtained by direct integration.

%
\section{Notation}
\la{se:notation}

Let us fix our notation here: we work with a semi-simple Lie algebra with hermitian generators $T^a$, whose real and  antisymmetric structure constants $f^{abc}$ are fixed by the commutation relation $[T^a,T^b]=i f^{abc}T^c$.
As usual, the quadratic Casimir operators of the fundamental (adjoint) representation (of dimensions $\NF$ and $\NA$, respectively)
are defined as $T^aT^a=\CF1\!\!1$ ($f^{acd}f^{bcd}=\CA\delta^{ab}$).
Furthermore, traces are normalized as ${\rm Tr}(T^aT^b)=\TF\delta^{ab}$, we denote the number of quark flavors with $\Nf$, and find it convenient to define the following normalized combinations:
\ba
\cf=\frac{\CF}{\CA} \quad,\quad \nf=\frac{\Nf\,\TF}{\CA} \;.
\ea

In our multi-loop diagrams, we will encounter traces of more than two group generators, giving rise to higher-order group invariants. 
It is useful to define traces over combinations of symmetric tensors \cite{vanRitbergen:1998pn},
of which we need the following (writing $[F^a]_{bc}=-if^{abc}$ for the generators of the adjoint representation):
\ba
d_1=\frac{[\str(T^aT^bT^cT^d)]^2}{\NA\TF^2\CA^2} \;,\;
d_2=\frac{\str(T^aT^bT^cT^d)\,\str(F^aF^bF^cF^d)}{\NA\TF\CA^3} \;,\;
d_3=\frac{[\str(F^aF^bF^cF^d)]^2}{\NA\CA^4} \;.
\ea
Here, $\str$ stands for a fully symmetrized trace (such that $\str(ABC)=\frac12{\rm Tr}(ABC+ACB)$ etc.).
While dealing with the quark sector, it might seem more natural to normalize these traces with respect to the dimension of the fundamental representation; to this end, we note the relation $\NA\TF=\NF\CF$ which holds in general \cite{vanRitbergen:1998pn,Georgi:1999wka}.
For the gauge group SU($N$) (where $\TF=\frac12$ and $\CA=N$), the normalized group invariants introduced above read \cite{vanRitbergen:1998pn}
\ba
\label{eq:sun}
\nf=\frac{\Nf}{2N}
\;,\quad
\cf=\frac{N^2-1}{2N^2}
\;,\quad
d_1=\frac{N^4-6N^2+18}{24N^4}
\;,\quad
d_2=\frac{N^2+6}{24N^2}
\;,\quad
d_3=\frac{N^2+36}{24N^2}\;.
\ea
The corresponding SU(3) values, relevant for physical QCD, hence read
\ba
\label{eq:su3}
\mbox{SU}(3):\quad
\nf=\frac{\Nf}{6}
\;,\quad
\cf=\frac{4}{9}
\;,\quad
d_1=\frac{5}{216}
\;,\quad
d_2=\frac{5}{72}
\;,\quad
d_3=\frac{5}{24}
\;.
\ea

%
\section{Quark mass renormalization}
\label{se:gm}

The renormalization constant for the quark mass $m_{\rm bare}=Z_m\,m_{\rm ren}$, or equivalently its anomalous dimension $\gamma_m = -\partial_{\ln\mu^2}\ln Z_m$, has been known at two \cite{Tarrach:1980up}
and three loops \cite{Tarasov:1982gk,Larin:1993tq} for a long time.
At four loops, $\gamma_m$  is known for SU(N) and QED \cite{Chetyrkin:1997dh} as well as for a general Lie group \cite{Vermaseren:1997fq}.
Presently, at five loops only the SU(3) value is publicly available \cite{Baikov:2014qja}. 
We present our corresponding result for a general Lie group below.

The structure of the quark mass anomalous dimension is 
\ba
\partial_{\ln\mu^2}\ln m_{\rm q}(\mu) &\equiv& \gamma_m(a) \;=\;
-\cf\,a\,\Big\{ 3 +\gamma_{m1}\,a +\gamma_{m2}\,a^2 +\gamma_{m3}\,a^3 +\gamma_{m4}\,a^4 +\dots \Big\}
\;,\\
\label{eq:adef}
a&\equiv&\frac{\CA\,g^2(\mu)}{16\pi^2}\;,
\ea
with $g(\mu)$ being the renormalized QCD gauge coupling constant that depends on the renormalization scale $\mu$
(we prefer to use the expansion parameter $a$ which is nothing but a rescaled version of the renormalized strong coupling constant $\alpha_s=\frac{g^2(\mu)}{4\pi}$). We work in $d=4-2\ep$ dimensions and employ the $\msbar$ scheme.
The coefficients $\gamma_{mn}$ are polynomials in $\nf$ and can be written in terms of our normalized 
group factors. Up to four loops, they read \cite{Vermaseren:1997fq}
\ba
3^1\,\gamma_{m1} &=& \nf\Big[-10\Big]+\Big[(9\cf+97)/2\Big] \;,\\
3^3\,\gamma_{m2} &=& \nf^2\Big[\!-\!140\Big] +\nf\Big[54(24\zeta_3\!-\!23)\cf -4(139\!+\!324\zeta_3)\Big]+\Big[(6966\cf^2\!-\!3483\cf\!+\!11413)/4\Big],\\
3^4\,\gamma_{m3} &=& \nf^3\Big[-8(83-144\zeta_3)\Big] +\nf^2\Big[48(19-270\zeta_3+162\zeta_4)\cf+2(671+6480\zeta_3-3888\zeta_4)\Big] 
\nonumber\\&&
+\nf\Big[-216(35-207\zeta_3+180\zeta_5)\cf^2-3(8819-9936\zeta_3+7128\zeta_4-2160\zeta_5)\cf
\nonumber\\&&
-(65459/2+72468\zeta_3-21384\zeta_4-32400\zeta_5)+2592(2-15\zeta_3)d_1\Big]
\nonumber\\&&
+\tfrac98\Big[-9(1261+2688\zeta_3)\cf^3+6(15349+3792\zeta_3)\cf^2-2(34045+5472\zeta_3-15840\zeta_5)\cf
\nonumber\\&&
+(70055+11344\zeta_3-31680\zeta_5)-1152(2-15\zeta_3)d_2\Big] \;,
\label{eq:g3}
\ea
where we have denoted values of the Riemann Zeta function as $\zeta_s=\zeta(s)=\sum_{n>0}n^{-s}$.

At five loops, from \app\ref{app:gracey}, we have LO and NLO large-$\Nf$ terms to all orders, coinciding with the leading terms above, and predicting the first two terms of the 5-loop contributions as
\ba
\label{eq:largeNf}
6^5\,\gamma_{m4} &=& \gamma_{m44}\,\Big[4\nf\Big]^4
+\gamma_{m43}\,\Big[4\nf\Big]^3+\gamma_{m42}\,\Big[4\nf\Big]^2+\gamma_{m41}\,\Big[4\nf\Big]+\gamma_{m40}\;,\\
\label{eq:gm44}
\gamma_{m44} &=& -6(65+80\zeta_3-144\zeta_4)\;,\\
\label{eq:gm43}
\gamma_{m43} &=& 3(4483\!+\!4752\zeta_3\!-\!12960\zeta_4\!+\!6912\zeta_5)\cf 
+(18667/2\!+\!32208\zeta_3\!+\!29376\zeta_4\!-\!55296\zeta_5)\;.
\ea
We have evaluated the remaining coefficients, mapping all diagrams onto fully massive vacuum tadpoles; IBP-reducing them to master integrals; using high-precision numerical evaluations thereof plus some additional consistency conditions to fix linear combinations of 12-line master integrals as explained in \se\ref{se:setup};
and finally employing PSLQ at 200 digits for discovery, and at 250 digits for confirmation, 
we confirm the two large-$\Nf$ expressions in \eqs\nr{eq:gm44} and \nr{eq:gm43}, and obtain the three missing coefficients of \eq\nr{eq:largeNf} in analytic form, containing only Zeta values\footnote{To make the group structure more visible, we resort to a vector notation here and below, where
a dot between two curly brackets denotes a scalar product as e.g.\ in $\{\cf,1\}.\{a,b\}=\cf a+b$.}:
\ba
\gamma_{m42} &=& \Big\{\cf^2,\cf,d_1,1\Big\}.\Big\{
9 (45253-230496 \zeta_{3}+48384 \zeta_{3}^2+70416 \zeta_{4}+144000 \zeta_{5}-86400 \zeta_{6}),
\nonumber\\&&
375373+323784 \zeta_{3}-1130112 \zeta_{3}^2+905904 \zeta_{4}-672192 \zeta_{5}+129600 \zeta_{6},
\nonumber\\&&
-864 (431 - 1371 \zeta_3 + 432 \zeta_4 + 420 \zeta_5),
\nonumber\\&&
4 (13709+394749 \zeta_{3}+173664 \zeta_{3}^2-379242 \zeta_{4}-119232 \zeta_{5}+162000 \zeta_{6})
\Big\}\;,
\\
\gamma_{m41} &=& \Big\{\cf^3,\cf^2,\cf d_1,\cf,d_1,d_2,1\Big\}.\Big\{
-54 (48797-247968 \zeta_{3}+24192 \zeta_{4}+444000 \zeta_{5}-241920 \zeta_{7}),
\nonumber\\&&
-18 (406861+216156 \zeta_{3}-190080 \zeta_{3}^2+254880 \zeta_{4}-606960 \zeta_{5}-475200 \zeta_{6}+362880 \zeta_{7}),
\nonumber\\&&
-62208 (11+154 \zeta_{3}-370 \zeta_{5}),
\nonumber\\&&
753557+15593904 \zeta_{3}-3535488 \zeta_{3}^2-6271344 \zeta_{4}-17596224 \zeta_{5}+1425600 \zeta_{6}+1088640 \zeta_{7},
\nonumber\\&&
1728 (3173-6270 \zeta_{3}+1584 \zeta_{3}^2+2970 \zeta_{4}-13380 \zeta_{5}),
\nonumber\\&&
1728 (380 - 5595 \zeta_3 - 1584 \zeta_3^2 - 162 \zeta_4 + 1320 \zeta_5),
\\&&
-2 (4994047+11517108 \zeta_{3}-57024 \zeta_{3}^2-5931900 \zeta_{4}-15037272 \zeta_{5}+4989600 \zeta_{6}+3810240 \zeta_{7})
\Big\}\;,\nonumber
\\
\gamma_{m40} &=& \Big\{\cf^4,\cf^3,\cf^2,\cf d_2,\cf,d_2,d_3,1\Big\}.\Big\{
972 (50995+6784 \zeta_{3}+16640 \zeta_{5}),
\nonumber\\&&
-54 (2565029+1880640 \zeta_{3}-266112 \zeta_{4}-1420800 \zeta_{5}),
\nonumber\\&&
108 (2625197+1740528 \zeta_{3}-125136 \zeta_{4}-2379360 \zeta_{5}-665280 \zeta_{7}),
\nonumber\\&&
373248 (141+80 \zeta_{3}-530 \zeta_{5}),
\nonumber\\&&
-8 (25256617+16408008 \zeta_{3}+627264 \zeta_{3}^2-812592 \zeta_{4}-40411440 \zeta_{5}+3920400 \zeta_{6}-5987520 \zeta_{7}),
\nonumber\\&&
-6912 (9598+453 \zeta_{3}+4356 \zeta_{3}^2+1485 \zeta_{4}-26100 \zeta_{5}-1386 \zeta_{7}),
\nonumber\\&&
5184 (537 + 2494 \zeta_3 + 5808 \zeta_3^2 + 396 \zeta_4 - 7820 \zeta_5 - 1848 \zeta_7),
\\&&
4 (22663417+\!10054464 \zeta_{3}+\!1254528 \zeta_{3}^2-\!1695276 \zeta_{4}-\!41734440 \zeta_{5}+\!7840800 \zeta_{6}+\!5987520 \zeta_{7})
\Big\}\;.\nonumber
\ea
Let us now discuss some checks on this new result.
The authors of \cite{Baikov:2014qja} have published the full 5-loop result for the case of SU(3).
To compare, we recall the definition of our expansion parameter in \eq\nr{eq:adef} and put all group invariants to their SU(3) values as given in \eq\nr{eq:su3}; the $\gamma_{m1..4}$ as listed above then coincide with the expressions given in \cite{Baikov:2014qja}.
Furthermore, the same group has very recently generalized their work to a general Lie 
group as well \cite{KostjaGammaM}. 
We have cross-checked their preliminary result with our $\gamma_{m4}$ as given above, and found full agreement. Since both five-loop results have been obtained with completely different methods (with the exception of also relying on \code{qgraf} for diagram generation, in \cite{KostjaGammaM} the 5-loop renormalization constants are mapped onto massless 4-loop two-point functions \cite{Smirnov:2010hd,Baikov:2010hf,Lee:2011jt}, and integral reduction is done via $1/d$ expansions \cite{Baikov:1996rk,Baikov:2005nv}), this agreement constitutes an extremely strong check.

%
\section{Quark field renormalization}
\label{se:g2}

For completeness, let us also present our new five-loop result for the quark field anomalous dimension 
$\gamma_2 = -\partial_{\ln\mu^2}\ln Z_2$, where the renormalization constant $Z_2$ relates bare and renormalized quark fields as 
$\psi_{\rm bare}=\sqrt{Z_2}\psi_{\rm ren}$.
As opposed to the quark mass, this quantity is not physical and hence gauge dependent.
Lower-loop results can be found for 
SU(N) and covariant gauge in \cite{Chetyrkin:1999pq},
and for a general Lie group with $\xi^0,\xi^1$ terms (which corresponds to a NLO expansion around Feynman gauge) in \cite{Czakon:2004bu}.
At five loops, $\gamma_2$ is known for SU(3) in Feynman gauge \cite{Baikov:2014qja},
and we will below once more present the generalization to a general Lie group.

Up to four loops, we obtain ($\xi$ being the covariant gauge parameter, such that the values $\xi=0/1$ correspond to Feynman/Landau gauge)
\ba
\gammaQQ &=& 
-\cf\,a\,\Big\{(1-\xi)+\gamma_{21}\,a+\gamma_{22}\,a^2+\gamma_{23}\,a^3+\gamma_{24}\,a^4+\dots\Big\}\;,\\
\label{eq:g21}
2^2\,\gamma_{21} &=& \nf\Big[-8\Big]+\Big[-6\cf+34-10\xi + \xi^2\Big]\;,\\
2^53^2\,\gamma_{22} &=& \nf^2\Big[640\Big]+ \nf\Big[8 (108 \cf - 1301 + 153 \xi)\Big] \\&& + \Big[432 \cf^2 -\! 72 (143 \!-\!48 \zeta_3)\cf +\!2 (10559 -\!1080 \zeta_3) -\!9 \xi (371 +\!48 \zeta_3) +\!27 \xi^2 (23 +\!4 \zeta_3) -\!90 \xi^3 \Big]\;,\nonumber\\
\label{eq:g23}
2^43^5\,\gamma_{23} &=& 
 \nf^3 \Big[ 13440 \Big]
+ \nf^2 \Big[ 6912 (19 - 18 \zeta_3)\cf + 16 (6835 + 9072 \zeta_3) + 64 \xi (269 - 324 \zeta_3) \Big]
\nonumber\\&&
+ \nf \Big[
5184 (19 - 48 \zeta_3) \cf^2 
+ \big(-108 (2407 - 1584 \zeta_3 - 1296 \zeta_4 - 5760 \zeta_5) 
\nonumber\\&&
+ 324 \xi (767 - 528 \zeta_3 - 144 \zeta_4) \big)\cf
+ 497664 d_1
-(1365691 + 154224 \zeta_3 + 97200 \zeta_4 + 311040 \zeta_5)
\nonumber\\&&
+ \xi (48865 + 152928 \zeta_3 + 29160 \zeta_4) 
- 54 \xi^2 (109 + 84 \zeta_3 - 18 \zeta_4)
\Big]
\nonumber\\&&
+\Big[
- 486 (1027 + 3200 \zeta_3 - 5120 \zeta_5) \cf^3 
+ 324 (5131 + 10176 \zeta_3 - 17280 \zeta_5) \cf^2 
\nonumber\\&&
+ \big(-108 (23777 + 7704 \zeta_3 + 2376 \zeta_4 - 28440 \zeta_5) - 1944 \xi (6 - 7 \zeta_3 + 10 \zeta_5)\big) \cf
\nonumber\\&&
+ 486 \big(16 (-33 + 95 \zeta_3 - 85 \zeta_5) - 8 \xi (1 + 48 \zeta_3 - 70 \zeta_5)
- 8 \xi^2 (7\zeta_3+5\zeta_5)
+ 20 \xi^3 (2\zeta_3-\zeta_5)
\nonumber\\&&
- \xi^4 (7\zeta_3-5\zeta_5)
\big) d_2
+ (10059589/4 - 241218 \zeta_3 + 168156 \zeta_4 - 604260 \zeta_5) 
\nonumber\\&&
- \xi (2127929/8 + 164106 \zeta_3 - 21141 \zeta_4 - 107730 \zeta_5)
+ 27 \xi^2 (13883 + 9108 \zeta_3 - 1548 \zeta_4 
\nonumber\\&&
- 1920 \zeta_5)/8
- 81 \xi^3 (263 + 65 \zeta_3 - 9 \zeta_4 + 20 \zeta_5)/2
+ 81 \xi^4 (57 + \zeta_3 + 10 \zeta_5)/4
\Big] \;.
\ea
We have presented the full gauge parameter dependence above. Note that this fills a gap in the literature and constitutes new information at four loops: in \cite{Czakon:2004bu} only the terms linear in $\xi$ have been evaluated for a general Lie group, while with full gauge dependence only the SU(N) result is available \cite{Chetyrkin:1999pq}. However, due to the degeneracies $2d_1=6\cf^2-5\cf+13/12$ and $2d_2=7/12-\cf$ in the SU(N) limit, one cannot uniquely extract the Lie group structure from the latter reference. Needless to say that our result for $\gamma_{23}$ given in \eq\nr{eq:g23} reproduces the $\xi^0$ and $\xi^1$ terms given in \cite{Czakon:2004bu}, and in the SU(N) limit reduces to the respective expressions of \cite{Chetyrkin:1999pq}, for all powers of $\xi$.

Expanding the all-order large-\/$\Nf$ Landau gauge result of \eq\nr{eq:g2NLO} in the coupling $\af$ allows to confirm \eqs\nr{eq:g21}-\nr{eq:g23} to NLO in $\nf$, and to predict the first two terms of $\gamma_{24}$ in that gauge as
\ba 
\label{eq:g24landau}
24^3\,\gamma_{24}|_{\xi=1}&=&
\frac{83-144\zeta_3}{72}\,\Big[16\nf\Big]^4
+\gamma_{243}^{\,\xi=1}\,\Big[16\nf\Big]^3
+\dots \;,\\
\gamma_{243}^{\,\xi=1} &=& \Big\{\cf, 1\Big\}.\Big\{ -659/18 + 312 \zeta_3 - 216 \zeta_4, -1783/36 - 248 \zeta_3 + 216 \zeta_4 \Big\} \;.
\ea

At five loops, along the same steps as explained in \se\ref{se:gm}, we have obtained the new Feynman gauge result\footnote{The restriction to $\xi=0$ is for practical reasons only. To evaluate the $\xi$\/-dependent coefficients, one would need to enlarge the integral reduction tables as produced by \code{crusher} and \code{TIDE} to integrals with higher propagator powers (or dots), roughly one more dot per power of the gauge parameter. Since the present calculation is at the limit of what the computing resources available to us are able to handle, we defer this to future work.}
\ba
24^3\,\gamma_{24} &=& 
\frac{83-144\zeta_3}{72}\,\Big[16\nf\Big]^4
+\gamma_{243}\,\Big[16\nf\Big]^3
+\gamma_{242}\,\Big[16\nf\Big]^2
+\gamma_{241}\,\Big[16\nf\Big]
+\gamma_{240}
+\order{\xi}\;,
\ea
where the coefficients again contain only Zeta values up to weight 7,
\ba
\label{eq:gamma243}
\gamma_{243} &=& \Big\{\cf, 1\Big\}.\Big\{ -659/18 + 312 \zeta_3 - 216 \zeta_4, -3443/48 - 255 \zeta_3 + 252 \zeta_4\Big\}\;,\\
\gamma_{242} &=& \Big\{\cf^2, \cf, d_1, 1\Big\}.\Big\{ -2 (2497 - 1200 \zeta_3 + 3456 \zeta_4 - 8640 \zeta_5), \nonumber\\&& 
 477433/12 - 45636 \zeta_3 + 4608 \zeta_3^2 + 11448 \zeta_4 - 65088 \zeta_5 + 28800 \zeta_6, 
 -384 (115 - 33 \zeta_3 - 90 \zeta_5), 
 \nonumber\\&&
 3015955/72 + 69509 \zeta_3 - 2304 \zeta_3^2 - 12861 \zeta_4 + 16662 \zeta_5 - 14400 \zeta_6 - 11907 \zeta_7 \Big\}\;,\\
\gamma_{241} &=& \Big\{\cf^3, \cf^2, \cf d_1, \cf, d_1, d_2, 1\Big\}.\Big\{ 24 (29209 + 89984 \zeta_3 + 12288 \zeta_3^2 - 28800 \zeta_4 - 187520 \zeta_5 + 
   76800 \zeta_6), \nonumber\\&& -4 (296177 + 517020 \zeta_3 + 26784 \zeta_3^2 - 469908 \zeta_4 - 
   4104720 \zeta_5 + 1069200 \zeta_6 + 3011904 \zeta_7), \nonumber\\&& -2304 (748 + 4536 \zeta_3 - 
   1368 \zeta_3^2 - 6780 \zeta_5 + 3255 \zeta_7), \nonumber\\&& 8 (115334 - 37764 \zeta_3 - 
   123012 \zeta_3^2 - 49923 \zeta_4 - 1124556 \zeta_5 + 133650 \zeta_6 + 
   1519308 \zeta_7), \nonumber\\&& 192 (16732 + 39912 \zeta_3 - 10944 \zeta_3^2 - 72960 \zeta_5 + 
   36771 \zeta_7), \nonumber\\&& 
   96 (6158 - 13952 \zeta_3 - 372 \zeta_3^2 + 2880 \zeta_4 - 39475 \zeta_5 - 3900 \zeta_6 + 45696 \zeta_7),
   \\&& -34919359/9 - 753797 \zeta_3 + 548148 \zeta_3^2 - 135063 \zeta_4 + 1759474 \zeta_5 + 
 265350 \zeta_6 - 2647806 \zeta_7\Big\}\;,\nonumber\\
\gamma_{240} &=& \Big\{\cf^4, \cf^3, \cf^2, \cf d_2, \cf, d_2, d_3, 1\Big\}.\Big\{ 1728 (4977 + 128000 \zeta_3 + 19968 \zeta_3^2 + 180800 \zeta_5 - 
   381024 \zeta_7), \nonumber\\&& -96 (835739 + 8494144 \zeta_3 + 1182336 \zeta_3^2 - 316800 \zeta_4 + 
   3983360 \zeta_5 + 844800 \zeta_6 - 17852688 \zeta_7), \nonumber\\&& 192 (825361 + 5472068 \zeta_3 + 
   651816 \zeta_3^2 - 335808 \zeta_4 - 1140420 \zeta_5 + 950400 \zeta_6 - 
   8056377 \zeta_7), \nonumber\\&& 4608 (10 + 53226 \zeta_3 - 15264 \zeta_3^2 + 2145 \zeta_5 - 
   45885 \zeta_7), \; -16 (84040774/9 \nonumber\\&& + 33396648 \zeta_3 + 2804616 \zeta_3^2 - 838782 \zeta_4 - 18160944 \zeta_5 + 
 6252300 \zeta_6 - 41015331 \zeta_7), \nonumber\\&& -384 (43066 + 628802 \zeta_3 - 160998 \zeta_3^2 + 36540 \zeta_4 - 201125 \zeta_5 - 
   53475 \zeta_6 - 403263 \zeta_7), \nonumber\\&& 
  -72 (20566 - 218812 \zeta_3 - 79080 \zeta_3^2 - 13212 \zeta_4 + 760220 \zeta_5 + 20100 \zeta_6 - 660667 \zeta_7),
   \; 804023630/9 \nonumber\\&& + 101490400 \zeta_3 + 3143352 \zeta_3^2 + 7356024 \zeta_4 - 
 86186276 \zeta_5 + 18372900 \zeta_6 - 115799439 \zeta_7 \Big\}\;.
\ea
As a speculation, comparing the Landau gauge prediction \eq\nr{eq:g24landau} with the Feynman gauge result \eq\nr{eq:gamma243}, the full result for $\gamma_{243}$ could be simply adding $\delta\gamma_{243}=\xi(3197/144+7\zeta_3-36\zeta_4)$; 
more generally however, consistency only requires that $\delta\gamma_{243}=\cf\,u_1(\xi)+u_2(\xi)$ with $u_1(0)=0=u_1(1)$ as well as $u_2(0)=0$ and $u_2(1)=3197/144+7\zeta_3-36\zeta_4$, the simplest
choice being the one speculated above.

As a check, replacing the group invariants with the values of \eq\nr{eq:su3} in our Feynman gauge result for $\gamma_{24}$, we find perfect agreement with the known SU(3) result of \cite{Baikov:2014qja} (see also \eq(46) of \cite{Baikov:2015tea}).

%
\section{Conclusions}
\la{se:conclu}

We have provided new results for two fundamental renormalization coefficients, at five loops and for
a semi-simple Lie group. In particular, our expression for the gauge-invariant quark mass anomalous dimension $\gamma_m$ coincides in various limits (large $\Nf$ as well as SU(3)) with previously known results,
and coincides exactly with recent results of another group \cite{KostjaGammaM}.
We have also provided the Feynman gauge result for the quark field anomalous dimension $\gamma_2$, which again could be checked against known expressions in the abovementioned limits. From these two quantities, one can reconstruct the two renormalization constants $Z_m$ and $Z_2$ of the quark sector, an electronic version of which is available by downloading the source of this article from {\tt http://arXiv.org/abs/1612.05512}.

As had already been observed in \cite{Vermaseren:1997fq}, looking at e.g.\ the 4-loop result for the quark mass anomalous dimension \eq\nr{eq:g3}, all Zeta terms (and also the higher group invariants $d_n$) vanish at $\{\cf=1,\nf=\frac12,d_1=d_2\}$, which  corresponds to ${\cal N}=1$ supersymmetry. The same had happened for the 4-loop Beta function (generalizing the last condition to $d_1=d_2=d_3$). 
For these parameters values, from \se\ref{se:gm} we have
\ba
\gamma_m &=& -a\Big\{3+\!16\,a+\!\tfrac{310}3\,a^2+\!\tfrac{2228}3\,a^3
+\!\Big(\tfrac{671075}{108} -\!194\,d_1
+\!\big(\tfrac{1483}{2} -\!2028\,d_1\big) \zeta_3 
-\!20\big(55 +\!354\,d_1\big) \zeta_5\Big) a^4\Big\} \;,\nonumber
\ea
where we observe that the cancellation pattern does not hold through five loops -- although the structure becomes much simpler, due to cancellation of all terms containing $\{\zeta_3^2,\zeta_4,\zeta_6,\zeta_7\}$.

To conclude the renormalization program at five loops, one needs to determine three more renormalization constants, which can be chosen to be those of the gluon and ghost fields, $Z_3$ and $Z_3^{c}$, respectively, and of the ghost-gluon vertex $Z_1^{ccg}$. From these, due to gauge invariance, one can then construct the renormalization constant for the gauge coupling (aka the Beta function, whose five-loop coefficient is so far known for SU(3) only \cite{Baikov:2016tgj}) as well as those for the remaining vertices. While the same methods that we have used here are sufficient to calculate those missing coefficients (which indeed already led to the NLO terms at large $\Nf$ \cite{Luthe:2016ima}), due to the complexity of the determination of $Z_3$ we leave the evaluation of the full coefficients for future work. 

%
\acknowledgments

We are indebted to K.~Chetyrkin for sending us their new five-loop results for $\gamma_m$ prior to publication \cite{KostjaGammaM}, to enable the important check discussed at the end of \se\ref{se:gm}.
The work of T.L.\ has been supported in part by DFG grants GRK 881 and SCHR 993/2.
A.M.\ is supported by a European Union COFUND/Durham Junior Research Fellowship under EU grant agreement number 267209.
P.M.\ was supported in part by the EU Network HIGGSTOOLS PITN-GA-2012-316704.
Y.S.\ acknowledges support from FONDECYT project 1151281 and UBB project GI-152609/VC.
All diagrams were drawn with Axodraw~\cite{Vermaseren:1994je,Collins:2016aya}.

%
\begin{appendix}
\section{Summary of large-$\Nf$ results}
\la{app:gracey}

Some coefficients of QCD anomalous dimensions are known to all loop orders, from a large $\Nf$ expansion. 
Taking $\nf$ and $\cf$ as above and defining 
\ba
\af\equiv\frac{\Nf \TF g^2(\mu)}{12\pi^2}\;=\;\frac{4\,\nf\,a}{3}
\quad,\qquad
\eta(\ep)\equiv\frac{(2\ep-3)\Gamma(4-2\ep)}{16\Gamma^2(2-\ep)\Gamma(3-\ep)\Gamma(\ep)}
\;,
\ea
the all-order leading-$\Nf$ \cite{Gracey:1993ua} and next-to-leading-$\Nf$ \cite{Ciuchini:1999cv,Ciuchini:1999wy} expressions that we need here read
\ba
\gamma_m&=&\frac{4\cf}\nf\Big\{\eta(\af)+\frac{\eta_3(\af)}{8\,\nf}+{\cal O}\Big(\frac1{\nf^2}\Big)\Big\}\;,\\
\label{eq:g2NLO}
\gammaQQ|_{\xi=1}&=&-\frac{2\af\cf}\nf\Big\{\eta(\af)+\frac1\nf\frac{\eta_4(\af)}{4\af}
+{\cal O}\Big(\frac1{\nf^2}\Big)\Big\}\;,
\ea
where the fact that the asymptotic expansions have been performed in Landau gauge only does not affect the physical and gauge invariant quark mass anomalous dimension $\gamma_m$.
To define the coefficient functions $\eta_3$ and $\eta_4$, it is convenient to define the linear combinations
\ba
\Psi(\ep) &=& \PolyGamma_0(1 - 2 \ep) + \PolyGamma_0(1 + \ep) - 
   \PolyGamma_0(1 - \ep) - \PolyGamma_0(1) \;,\\
\Phi(\ep) &=& \PolyGamma_1(1 - 2 \ep) - \PolyGamma_1(1 + \ep) - 
   \PolyGamma_1(1 - \ep) + \PolyGamma_1(1) \;,\\
\Theta(\ep) &=& \PolyGamma_1(1 - \ep) - \PolyGamma_1(1) \;,
\ea
where $\PolyGamma_n(x)=\partial_x^{n+1}\ln\Gamma(x)$ is the PolyGamma function.
Then \cite{Ciuchini:1999cv,Ciuchini:1999wy}
\ba
\eta_3(\ep) &\equiv&
\Big(\!-\!\frac{11}4 +\! \sum_{n>0}\frac{f_n\ep^n}n\Big) 8 \ep \partial_\ep\eta(\ep) 
-\frac{16 \eta^2(\ep)}{(3 \!-\! 2 \ep)(1 \!-\! \ep)} 
\Big\{3 (2 - \ep)^2 (1 - \ep)^2 \Theta(\ep)-(5 + 5 \ep - 11 \ep^2 + 4 \ep^3) , 
\nonumber\\&&
(88 \!-\! 372 \ep \!+\! 551 \ep^2 \!-\! 380 \ep^3 \!+\! 160 \ep^4 \!-\! 64 \ep^5 \!+\! 16 \ep^6)
\!-\! 4 \ep (3 \!-\! 2 \ep) (1 \!-\! 2 \ep) (2 \!-\! \ep) (1 \!-\! \ep)^2 \big(\Psi^2(\ep) \!+\! \Phi(\ep)\big) 
\nonumber\\&&
+ 2 (1 - \ep) (24 - 144 \ep + 249 \ep^2 - 146 \ep^3 + 12 \ep^4 + 8 \ep^5) \Psi(\ep)\Big\} . 
\Big\{\tfrac{2}{2 - \ep}\,\cf, \tfrac1{4\ep(3 - 2 \ep)(1 - 2 \ep)}\Big\} \;,\\
\eta_4(\ep) &=& \frac{\ep\eta_3(\ep)}2+\Big(\!-\!\frac{11}4 +\! \sum_{n>0}\frac{f_n\ep^n}n\Big) 4\ep\eta(\ep)
+\frac{2\eta^2(\ep)}{3-2\ep} \Big\{-8(1-4\ep+2\ep^2),\tfrac{(2-5\ep+2\ep^2)^2}{1-\ep}\Big\} . \Big\{\cf,1\Big\} \;,\\
{\rm with}&&\sum_{j>0}f_j\,\ep^j \;\equiv\; -\eta(\ep)
\Big\{4(1+\ep)(1-2\ep)\cf+\tfrac{4\ep^4-14\ep^3+32\ep^2-43\ep+20}{(1-\ep)(3-2\ep)}\Big\} \;.
\ea
The coefficients $\af$,  $\eta(\ep)$ and $f_j$ are the same as we had defined in \cite{Luthe:2016ima}.

\end{appendix}

%


\begin{thebibliography}{99}

\bibitem{Baikov:2005rw}
  P.~A.~Baikov, K.~G.~Chetyrkin and J.~H.~K\"uhn,
  {\em Scalar correlator at O($\alpha_s^4$), Higgs decay into b-quarks and bounds on the light quark masses},
  Phys.\ Rev.\ Lett.\  {\bf 96} (2006) 012003
  [hep-ph/0511063].

\bibitem{Tarrach:1980up}
  R.~Tarrach,
  {\em The Pole Mass in Perturbative QCD},
  Nucl.\ Phys.\ B {\bf 183} (1981) 384.

\bibitem{Tarasov:1982gk}
  O.~V.~Tarasov,
  {\em Anomalous Dimensions Of Quark Masses In Three Loop Approximation},
  JINR-P2-82-900 (in Russian).

\bibitem{Larin:1993tq}
  S.~A.~Larin,
  {\em The Renormalization of the axial anomaly in dimensional regularization},
  Phys.\ Lett.\ B {\bf 303} (1993) 113
  [hep-ph/9302240].

\bibitem{Vermaseren:1997fq}
  J.~A.~M.~Vermaseren, S.~A.~Larin and T.~van Ritbergen,
  {\em The four loop quark mass anomalous dimension and the invariant quark mass},
  Phys.\ Lett.\ B {\bf 405} (1997) 327
  [hep-ph/9703284].

\bibitem{Chetyrkin:1997dh}
  K.~G.~Chetyrkin,
  {\em Quark mass anomalous dimension to O($\alpha_s^4$)},
  Phys.\ Lett.\ B {\bf 404} (1997) 161
  [hep-ph/9703278].

\bibitem{Baikov:2014qja}
  P.~A.~Baikov, K.~G.~Chetyrkin and J.~H.~K\"uhn,
  {\em Quark Mass and Field Anomalous Dimensions to ${\cal O}(\alpha_s^5)$},
  JHEP {\bf 1410} (2014) 076
  [arXiv:1402.6611].

\bibitem{Luthe:2016ima}
  T.~Luthe, A.~Maier, P.~Marquard and Y.~Schr\"oder,
  {\em Towards the five-loop Beta function for a general gauge group},
  JHEP {\bf 1607} (2016) 127
  [arXiv:1606.08662].

\bibitem{Nogueira:1991ex}
  P.~Nogueira,
  {\em Automatic Feynman graph generation},
  J.\ Comput.\ Phys.\  {\bf 105} (1993) 279;

\bibitem{Nogueira:2006pq}
  P.~Nogueira,
  {\em Abusing qgraf},
  Nucl.\ Instrum.\ Meth.\ A {\bf 559} (2006) 220.

\bibitem{Vermaseren:2000nd}
  J.~A.~M.~Vermaseren,
  {\em New features of FORM},
  math-ph/0010025.

\bibitem{Tentyukov:2007mu}
  M.~Tentyukov and J.~A.~M.~Vermaseren,
  {\em The Multithreaded version of FORM},
  Comput.\ Phys.\ Commun.\ {\bf 181} (2010) 1419
  [hep-ph/0702279].

\bibitem{Kuipers:2012rf}
  J.~Kuipers, T.~Ueda, J.~A.~M.~Vermaseren and J.~Vollinga,
  {\em FORM version 4.0},
  Comput.\ Phys.\ Commun.\  {\bf 184} (2013) 1453
  [arXiv:1203.6543].

\bibitem{vanRitbergen:1998pn}
  T.~van Ritbergen, A.~N.~Schellekens and J.~A.~M.~Vermaseren,
  {\em Group theory factors for Feynman diagrams},
  Int.\ J.\ Mod.\ Phys.\ A {\bf 14} (1999) 41
  [hep-ph/9802376].

\bibitem{Misiak:1994zw}
  M.~Misiak and M.~M\"unz,
  {\em Two loop mixing of dimension five flavor changing operators},
  Phys.\ Lett.\ B {\bf 344} (1995) 308
  [hep-ph/9409454].
  
\bibitem{vanRitbergen:1997va}
  T.~van Ritbergen, J.~A.~M.~Vermaseren and S.~A.~Larin,
  {\em The Four loop beta function in quantum chromodynamics},
  Phys.\ Lett.\ B {\bf 400} (1997) 379
  [hep-ph/9701390].

\bibitem{Chetyrkin:1997fm}
  K.~G.~Chetyrkin, M.~Misiak and M.~M\"unz,
  {\em Beta functions and anomalous dimensions up to three loops},
  Nucl.\ Phys.\ B {\bf 518} (1998) 473
  [hep-ph/9711266].

\bibitem{Luthe:2015ngq}
  T.~Luthe,
  {\em Fully massive vacuum integrals at 5 loops}, PhD thesis, Bielefeld University 2015.

\bibitem{crusher}
  P.~Marquard and D.~Seidel, \code{crusher} (unpublished).

\bibitem{Chetyrkin:1981qh}
  K.~G.~Chetyrkin and F.~V.~Tkachov,
  {\em Integration by Parts: The Algorithm to Calculate beta Functions in 4 Loops},
  Nucl.\ Phys.\ B {\bf 192} (1981) 159.

\bibitem{Laporta:2001dd}
  S.~Laporta,
  {\em High precision calculation of multiloop Feynman integrals by difference equations},
  Int.\ J.\ Mod.\ Phys.\ A {\bf 15} (2000) 5087
  [hep-ph/0102033].

\bibitem{Luthe:2016sya}
  T.~Luthe and Y.~Schr\"oder,
  {\em Fun with higher-loop Feynman diagrams},
  J.\ Phys.\ Conf.\ Ser.\  {\bf 762} (2016) no.1,  012066
  [arXiv:1604.01262].

\bibitem{fermat}
  R.~H.~Lewis, \code{fermat}, {\tt http://home.bway.net/lewis/}

\bibitem{MR1489971}
  H.~R.~P.~Ferguson, D.~H.~Bailey and S.~Arno, 
  {\em Analysis of PSLQ, an integer relation finding algorithm},
  Math.\ Comput.\ {\bf 68} (1999) 351.

\bibitem{Smirnov:2012gma}
  V.~A.~Smirnov,
  {\em Analytic tools for Feynman integrals},
  Springer Tracts Mod.\ Phys.\  {\bf 250} (2012) 1.

\bibitem{Roth:1996pd}
  M.~Roth and A.~Denner,
  {\em High-energy approximation of one loop Feynman integrals},
  Nucl.\ Phys.\ B {\bf 479} (1996) 495
  [hep-ph/9605420].

\bibitem{Binoth:2003ak}
  T.~Binoth and G.~Heinrich,
  {\em Numerical evaluation of multiloop integrals by sector decomposition},
  Nucl.\ Phys.\ B {\bf 680} (2004) 375
  [hep-ph/0305234].

\bibitem{Smirnov:2015mct}
  A.~V.~Smirnov,
  {\em FIESTA4: Optimized Feynman integral calculations with GPU support},
  Comput.\ Phys.\ Commun.\  {\bf 204} (2016) 189
  [arXiv:1511.03614].

\bibitem{Luthe:2016spi}
  T.~Luthe and Y.~Schr\"oder,
  {\em Five-loop massive tadpoles},
  PoS LL {\bf 2016} (2016) 074
  [arXiv:1609.06786].
    
\bibitem{inPrep}
  T.~Luthe, A.~Maier, P.~Marquard and Y.~Schr\"oder,
  in preparation.
  
\bibitem{Georgi:1999wka}
  H.~Georgi,
  {\em Lie algebras in particle physics},
  Front.\ Phys.\  {\bf 54} (1999) 1.

\bibitem{KostjaGammaM}
  P.~A.~Baikov, K.~G.~Chetyrkin and J.~H.~K\"uhn,
  private communication.

\bibitem{Smirnov:2010hd}
  A.~V.~Smirnov and M.~Tentyukov,
  {\em Four Loop Massless Propagators: a Numerical Evaluation of All Master Integrals},
  Nucl.\ Phys.\ B {\bf 837} (2010) 40
  [arXiv:1004.1149].

\bibitem{Baikov:2010hf}
  P.~A.~Baikov and K.~G.~Chetyrkin,
  {\em Four Loop Massless Propagators: An Algebraic Evaluation of All Master Integrals},
  Nucl.\ Phys.\ B {\bf 837} (2010) 186
  [arXiv:1004.1153].

\bibitem{Lee:2011jt}
  R.~N.~Lee, A.~V.~Smirnov and V.~A.~Smirnov,
  {\em Master Integrals for Four-Loop Massless Propagators up to Transcendentality Weight Twelve},
  Nucl.\ Phys.\ B {\bf 856} (2012) 95
  [arXiv:1108.0732].

\bibitem{Baikov:1996rk}
  P.~A.~Baikov,
  {\em Explicit solutions of the three loop vacuum integral recurrence relations},
  Phys.\ Lett.\ B {\bf 385} (1996) 404
  [hep-ph/9603267].

\bibitem{Baikov:2005nv}
  P.~A.~Baikov,
  {\em A Practical criterion of irreducibility of multi-loop Feynman integrals},
  Phys.\ Lett.\ B {\bf 634} (2006) 325
  [hep-ph/0507053].

\bibitem{Chetyrkin:1999pq}
  K.~G.~Chetyrkin and A.~Retey,
  {\em Renormalization and running of quark mass and field in the regularization invariant and MS-bar schemes at three loops and four loops},
  Nucl.\ Phys.\ B {\bf 583} (2000) 3
  [hep-ph/9910332].
  
\bibitem{Czakon:2004bu}
  M.~Czakon,
  {\em The Four-loop QCD beta-function and anomalous dimensions},
  Nucl.\ Phys.\ B {\bf 710} (2005) 485
  [hep-ph/0411261].

\bibitem{Baikov:2015tea}
  P.~A.~Baikov, K.~G.~Chetyrkin and J.~H.~K\"uhn,
  {\em Massless Propagators, $R(s)$ and Multiloop QCD},
  Nucl.\ Part.\ Phys.\ Proc.\  {\bf 261-262} (2015) 3
  [arXiv:1501.06739].
  
\bibitem{Baikov:2016tgj}
  P.~A.~Baikov, K.~G.~Chetyrkin and J.~H.~K\"uhn,
  {\em Five-Loop Running of the QCD coupling constant},
  arXiv:1606.08659.

\bibitem{Vermaseren:1994je}
  J.~A.~M.~Vermaseren,
  {\em Axodraw},
  Comput.\ Phys.\ Commun.\  {\bf 83} (1994) 45.

\bibitem{Collins:2016aya}
  J.~C.~Collins and J.~A.~M.~Vermaseren,
  {\em Axodraw Version 2},
  arXiv:1606.01177.
  
\bibitem{Gracey:1993ua}
  J.~A.~Gracey,
  {\em Quark, gluon and ghost anomalous dimensions at O(1/$N_f$) in quantum chromodynamics},
  Phys.\ Lett.\ B {\bf 318} (1993) 177
  [hep-th/9310063].

\bibitem{Ciuchini:1999cv}
  M.~Ciuchini, S.~E.~Derkachov, J.~A.~Gracey and A.~N.~Manashov,
  {\em Quark mass anomalous dimension at O(1/$N_f^2$) in QCD},
  Phys.\ Lett.\ B {\bf 458} (1999) 117
  [hep-ph/9903410].

\bibitem{Ciuchini:1999wy}
  M.~Ciuchini, S.~E.~Derkachov, J.~A.~Gracey and A.~N.~Manashov,
  {\em Computation of quark mass anomalous dimension at O(1/$N_f^2$) in quantum chromodynamics},
  Nucl.\ Phys.\ B {\bf 579} (2000) 56
  [hep-ph/9912221].

\end{thebibliography}
\end{document}